# The Grammar of FAIR

*A Granular Architecture of Semantic Units for FAIR Semantics, Inspired by Biology and Linguistics*

Vogt, Lars[1]; Mons, Barend[2];

[1] *TIB Leibniz Information Centre for Science and Technology, Welfengarten 1B, 30167 Hanover, Germany,*  orcid.org/0000-0002-8280-0487

[2] *Leiden Initiative for FAIR and Equitable Science, Rijnsburgerweg 10, 2333 AA Leiden, The Netherlands*  ORCID.org/0000-0003-3934-0072

Correspondence to: lars.m.vogt@googlemail.com

# Abstract

The FAIR Principles aim to make data and knowledge Findable, Accessible, Interoperable, and Reusable, yet current digital infrastructures often lack a unifying semantic framework that bridges human cognition and machine-actionability. In this paper, we introduce the *Grammar of FAIR*: a granular and modular architecture for FAIR semantics built on the concept of semantic units. Semantic units, comprising atomic statement units and composite compound units, implement the principle of semantic modularisation, decomposing data and knowledge into independently identifiable, semantically meaningful, and machine-actionable units. A central metaphor guiding our approach is the analogy between the hierarchy of level of organisation in biological systems and the hierarchy of levels of organisation in information systems: both are structured by granular building blocks that mediate across multiple perspectives while preserving functional unity. Drawing further inspiration from concept formation and natural language grammar, we show how these building blocks map to FAIR Digitial Objects (FDOs), enabling format-agnostic semantic transitivity from natural language token models to schema-based representations. This dual biological-linguistic analogy provides a semantics-first foundation for evolving cross-ecosystem infrastructures, paving the way for the Internet of FAIR Data and Services (IFDS) and a future of modular, AI-ready, and citation-granular scholarly communication.

# Introduction and Motivation

We are living in an era of rapidly accelerating digital knowledge production. The global scientific enterprise is expanding at an unprecedented rate: over seven million scholarly articles documenting **scientific knowledge** are published each year (1), alongside a growing volume of associated **research data**, software, and workflows deposited in domain-specific and general-purpose repositories (such as Zenodo, Figshare, Dryad, GitHub, and WDS). Critically, the central artefacts of scholarly communication, i.e., academic papers, remain largely designed for human consumption, limiting their utility in machine-assisted knowledge discovery, synthesis, and decision-making. At the same time, automation and artificial intelligence (AI) are becoming increasingly embedded in scientific workflows, offering new ways to navigate, process, analyse, and interpret the growing body of data, information, and knowledge, thereby fundamentally changing the way in which research is conducted. It is increasingly clear that human-machine interaction, utilising neural learning, symbolic representation, knowledge graphs, and ontologies to build collaborative AI that participates in hybrid teams of humans and AI assistants, is important for the responsible use of AI (hybrid intelligence), as '*AI on its own*' is prone to error (55-57).

While these developments hold considerable promise, they also present a profound epistemological and infrastructural challenge: **How can scientific data, information, and knowledge be presented in ways that are intelligible to both humans and machines, without reducing or obscuring their inherent complexity?** Addressing this challenge requires that we go beyond strategies for simply making data, information, knowledge, and software findable and accessible. It compels us to confront the deeper question of how meaning, i.e., semantics, can be structured and formalised for reliable (re)use across distributed, digital environments, without the necessity to involve the original creators of the data.

The **FAIR Guiding Principles** (**F**indable, **A**ccessible, **I**nteroperable, and **R**eusable) (2) have been instrumental in shaping global efforts to enhance machine actionability, thus the inherent quality and reuse of scientific data and knowledge. Initially framed to improve data stewardship, the principles were also recommended to be applied more broadly to a diverse range of digital research objects, including metadata, software, workflows, samples, and more. In this evolving landscape, digital objects are increasingly expected to be FAIR and thus not designed for processing only by humans, but also by machines. This shift has led to new infrastructures and frameworks designed to support the creation, dissemination, and integration of FAIR digital research objects. Among these, the concept of the **FAIR Digital Object (FDO)** (3,4, 54) has gained prominence as a generic abstraction for making heterogeneous digital artefacts part of an interoperable, machine-actionable knowledge ecosystem, the **Internet of FAIR Data and Services (IFDS)** (5).

However, realising FAIRness is not solely a technical endeavour. It is simultaneously epistemic, organisational, and social. The fundamental tension in modern scientific communication is this: humans and machines consume and process information in fundamentally different ways. Machines require formal, logically interlinked structures to support any form of simple reasoning, pattern recognition, discovery, and automation. Humans, by contrast, require contextualised, cognitively interoperable, and narratively coherent information, and have a strongly developed ability of 'on the fly' disambiguation based on contextual information. Bridging that gap requires more than just better formats or richer metadata. It demands a fundamental rethinking of how we model, partition, and communicate data, information and knowledge, from atomic measurements to high-level theories. And a truly FAIR knowledge infrastructure must serve both needs: preserving the nuance and richness of human communication while enabling the scalability and precision required by machines.

This dual imperative introduces specific design requirements.

Scientific data and knowledge must be:

- Machine-readable, -interpretable, and -actionable, adhering to the **FAIR Principles** (2); and
- Human-interpretable, cognitively interoperable, and semantically coherent, adhering to the **CLEAR Principle**. According to the CLEAR Principle (6), data and knowledge must be **C**ognitively interoperable, semantically **L**inked, contextually **E**xplorable, intuitively **A**ccessible, and human-**R**eadable and -interpretable. These requirements are operationalised through three criteria (see Box 1), each emphasising the need to structure data, information and knowledge into semantically meaningful modules and subsets, each identifiable, reference-able, and reusable across contexts.

| **Box 1 \| The CLEAR Principle with its three criteria** (6) |
|---|
| C1    (meta)data are structured into semantically meaningful statement subsets, each modelling an individual proposition that is represented by its own globally unique and persistent identifier (ideally forming a statement in the form of a statement FAIR Digital Object) that enables its independent actionability, its referencing, and its identification, and that instantiates a corresponding semantically defined (statement FAIR Digital Object) class |
| C2    (meta)data statements and compound subsets can be (recursively) combined to form compound subsets (i.e., nested FAIR Digital Objects), each with its own globally unique and persistent identifier, instantiating a corresponding semantically defined (nested FAIR Digital Object) class |
| C3    human-readable textual and graphical display of (meta)data is decoupled from machine-actionable (meta)data storage to reduce the complexity of (meta)data displayed to human readers and to display only information that is relevant to human readers, and a user interface supports contextual exploration of the (meta)data |

Despite progress in standardising identifiers, metadata schemata, and access protocols, a critical question remains insufficiently addressed: **What does it mean to make *meaning* FAIR?** How can the semantics, i.e., the conceptual referents of data and digital artefacts, be structured, referenced, and reused in a FAIR-compliant manner? To meet this challenge, we must also focus our attention to the **FAIR semantics** of the data, information and knowledge. This involves confronting complexity head-on, not by eliminating it, but by developing representational frameworks that help us manage it.

This paper approaches this challenge by focusing on the problem of complexity. We argue that **FAIR semantics** requires confronting the inherent **complexity of the world** and the **equally complex ways in which humans observe, understand/model, and communicate about it**. Scientific knowledge is complex in several senses: it is abstract, multi-layered, theory-laden, context-sensitive, and often contested or provisional. It evolves over time and spans multiple levels of granularity and generality. Researchers routinely engage in abstraction, generalisation, and interpretation, employing diverse representational strategies including formal models, conceptual schemata, narratives, visualisations, and diagrams. Supporting this diversity of meaning-making practices in digital infrastructures requires more than merely rendering data and knowledge FAIR and CLEAR. It requires recognising and enabling **hybrid reasoning practices**, where human understanding and machine-based operations converge.

The growing adoption of FDOs provides a timely opportunity to move beyond the syntactic and technical layer of FAIR implementation, toward addressing this deeper semantic dimension. To frame this additional layer, we introduce two metaphors that highlight common strategies for managing complexity and making it communicable. These metaphors align with our proposed principle of **semantic modularisation** (7).

The first metaphor derives from **natural language and concept formation**. Faced with complex or novel phenomena, humans routinely **coin new terms** to serve as labels for previously unnamed concepts, entities, relationships, or patterns. These terms allow us to refer to complex constructs as unified semantic wholes, such as "*ecosystem*," "*quark*," "*prion,*" or "*peer review*". Once coined, these terms become part of a community's conceptual repertoire. They enable **shared reference** to otherwise difficult-to-articulate phenomena and facilitate shared understanding, communication, and reasoning and thus **discourse *about*** their referents: they become **semantic handles** that can be integrated into reasoning, argumentation, and formal analysis. These novel terms are, however, not primitive or context-free, but embedded in theories, assumptions, and methodological practices.

This first metaphor aligns with the **semantic units** approach (6–8), which identifies **semantically meaningful subsets** within datasets, assigning each a **persistent identifier** that instantiates a corresponding **class of information**. This process of **semantic partitioning** reflects a broader strategy: to organise data into modular, intelligible structures that mirror the granular organisation of the referent system being modelled.

The second metaphor comes from **levels of organisation in the life sciences**. Biological systems are described across nested levels: atoms, molecules, cells, tissues, organs, organisms, populations, and ecosystems. Each level adds complexity and emergent properties, while preserving coherence with adjacent levels, introducing new organisational features and explanatory frameworks. Researchers shift between these levels depending on context, preserving structured reasoning while accommodating scale-sensitive analysis.

This second metaphor underscores the importance of **granularity** and thus of being able to view and structure composite data, information, and knowledge at different scales while preserving the semantic integrity of the building blocks. It helps us think about how complex wholes can be analysed in terms of their related parts, and how scientific and biological discourse shifts fluidly between different levels of abstraction and explanation. It also maps onto the granular structure of semantic units, where units can be composed hierarchically, yielding a nested structure of semantic units within semantic units that accommodates abstraction, composition, and decomposition, resulting in a nested, granular organisation of a given dataset (6–8). This layered approach mimics the way in which biological systems work, retain integrity, and evolve.

Together, these metaphors point toward the necessity of **explicit, referenceable, and compositional representations of semantic content**. We propose that **semantic modularisation**, using semantic units as foundational building blocks (7), offers a viable **granular architecture for FAIR semantics**. Such an architecture would allow complex knowledge to be expressed, referenced, and reused at different levels of abstraction and composition. Semantic units, organised into multiple granular perspectives and implemented as FAIR Digital Objects, allow for scalable, semantically grounded, and perspectival structuring of complex knowledge domains, providing **referential clarity** while supporting **compositionality** and **contextualisation**. This approach does not rely on a single unified and static ontology, but supports partial, evolving, and multi-level representations of meaning, reflecting a deeper logic of **semantic modularity** that enhances traceability, composability, knowledge evolution, and reuse in distributed environments.

Taken together, these components form what we call the **Grammar of FAIR**: a semantically layered architecture for structuring scientific meaning in a way that is both machine-actionable and meaningful to humans. This grammar is not merely syntactic but represents a framework for extending the FAIR principles beyond data, toward a deeper and more robust representation of '*machine actionable meaning*'.

This paper develops a detailed proposal of how this might be achieved.

In *Data, Information, and Knowledge Representations as Models of Real-World Referent Systems*, we pick up the first metaphor and characterise data and knowledge as token and universal metamodels, and modelling as a partitioning activity in which the granular part in the foreground becomes distinguishable and thus referenceable. This leads us to *Modelling Granularity and the Granularity of Models*, in which we introduce a theory of granularity that provides the granulation criteria necessary for logically and ontologically sound partitioning and thus for model building. Here, we also discuss the second metaphor and propose a logically and ontologically consistent hierarchy of levels of organisation. In *Semantic Modularisation and the Concept of Semantic Units*, we introduce the semantic modularisation principle and the concept of semantic units, and discuss in *Semantic Units as Semantic FAIR Digital Objects and Information Granularity* that semantic FDO types must map to semantic unit types and that the publication of these semantic FDOs should be format-flexible, with [Nanopublications](#) and [Research Object Crates](#) (RO-Crates) as possible serialisations. We then discuss how information granularity can be represented with nested semantic FDOs and identify an information hierarchy that we derived from our knowledge about the levels of biological organisation. In *FAIR Semantics and the Granularity of FAIRness*, we discuss that the assessment of FAIRness should take into account information granularity, resulting in a granular FAIRness. With *Toward a Grammar of FAIR*, we want to stimulate a discussion on what infrastructures and FAIR and CLEAR services are required to support FAIR semantics, with natural language metamodels taking the role of a *lingua franca* for realising semantic coherence and transitive schema representations across FDOs. We conclude by discussing how FDOs that are based on semantic units can potentially reshape the architecture of scholarly communication.

# Data, Information, and Knowledge Representations as Models of Real-World Referent Systems

All information about the real world, whether in the form of data, information, or knowledge, can be regarded as a **model** of some **referent system** that exists within that world. A model, in general, represents information (i.e., meaning) about something; it is created by a sender for a receiver, with a specific **usage context** in mind, and serves as a **surrogate** for the referent system it represents (9).

Models can be characterised as being derived from a referent system that they attempt to represent (i.e., **mapping feature**), to be **abstractions** of this referent system as they model only a fraction of the referent's elements and properties (i.e., **reduction feature**), and to be usable as a substitute for the referent system so that its responses are consistent with those of the referent system in the context of the intended usage (i.e., **pragmatic feature**) (10,11).

Two foundational categories of models can be distinguished. **Token models** capture the properties and relationships between **individual entities** (i.e., particulars). **Empirical data**, whether represented in natural language or as data structures, are instances of token models (12), such as "*Parasite X has a mass of 24.76 grams.*"

**Type models**, in contrast, abstract over individual entities, and capture the properties and relationships between kinds or categories of entities (i.e., universals, types). Type models are often derived from token models through **classification** of their elements and properties (10) (e.g., "*PARASITE has a MASS of VALUE GRAM-BASED_UNIT*"). Thus, token models instantiate type models. A further refinement leads to **metamodels**, a subcategory of type models that **generalise** over elements from existing type models (e.g., "*MATERIAL_OBJECT has a QUALITY of VALUE UNIT*"). **Scientific knowledge** in the form of universally applicable theories, hypotheses, rules, and causal relationships represent **universal metamodels**, and **data schemata** represent **schema metamodels**. Together, token models, type models, and metamodels form a hierarchy of abstraction in how we present the world.

In scientific research, we ideally aim to model the granular structure of the real world. Our empirical data and scientific knowledge should reflect this structure, enabling accurate, scalable, and cognitively tractable representations. This is insofar important, as these models help us communicate our **cognitive representations** (mental models) of **real-world entities** (objects, processes, and their properties and relations) in the form of token, type, and metamodels by translating them into **textual representational artefacts** (text, diagrams, and data structures) and thus symbolic representations of these models (for *real entities*, *cognitive representations*, and *textual representational artefacts*, see (12–14)). In this process, data and knowledge are translated from mental to digital or physical representations with the goal of enabling sharing understanding between sender and receiver by sharing their cognitive representations.

Modelling always involves **partitioning**. Some element, property, or relation is brought into focus by placing it in the foreground and becomes a **distinguishable and referenceable part** of the referent system. Other parts recede into the background, but may still play a role, consciously or unconsciously. This act of partitioning mirrors the **natural language metaphor** introduced earlier: just as we coin new terms to name complex, previously unnamed phenomena, we construct models that make parts of the world **explicitly referenceable**. Actually, we first have to build these models before we can coin new terms to linguistically refer to them.

Modelling is therefore **cognitive**, **ontological**, and **representational** and thus **pre-linguistic**. Across disciplines, humans construct and navigate complex real-world systems by abstracting and grouping their elements, often **coining new terms** based on partitioning the world into some granular part in the foreground and everything else in the background, to describe relationships and phenomena that were previously ineffable. Such semantic coinage allows us to *talk about* higher-order relationships and ideas that would otherwise be too complex to handle.

Importantly, partitioned elements can themselves be partitioned further, leading to nested **granular partitions**. Given the granular nature and ontological richness of real-world systems, along with our varied cognitive preconditions, cognitive interests, and anticipated usage contexts, modelling inevitably produces **alternative partitions** and thus multiple models of the same referent system. These may reflect different perspectives, but under logical and ontological alignment, they can often be reconciled into integrated models.

Thus, the act of modelling is inherently **intentional**, **situated**, and **perspectival**. To yield reliable, reusable, and logically consistent results, modelling must be governed by **sound logical and ontological principles**.

# Modelling Granularity and the Granularity of Models

Granular partitions reflect structured hierarchies, such as the various **hierarchies of levels of organisation** published in the life sciences, which are used to model increasing complexity. These hierarchies mirror the structure of the world as composed of structurally and functionally stable entities, i.e., **building blocks**, forming basic types of material entities that can be recursively combined to form more complex systems. Such structures are critical not only for understanding natural complexity and how this complexity supports life and could have evolved, but also for modelling it, particularly in the digital realm, where machine-actionable, logically consistent representations are also essential.

Before we turn to the levels of organisation metaphor introduced earlier, we must critically assess how granular hierarchies are typically constructed. Many existing hierarchies, though intuitively compelling, lack logical and ontological consistency. This makes them difficult to model formally and limits their utility for machine interpretation.

A prominent example is **Eldredge's somatic hierarchy** (15), which progresses from atoms to ecosystems: *atoms > molecules > cells > tissues > organs > organ systems > organisms > populations > ecosystems*. However, this hierarchy conflates different **granulation criteria**, shifting between spatio-structural criteria (*atoms*, *molecules*, *cells*), functional criteria (*organs*, *organ systems*, *organisms*, *ecosystems*), and scale and resolution criteria (*tissues*, *populations*), leading to category inconsistencies. For example, single-celled organisms may occupy multiple levels simultaneously, undermining the coherence and machine-actionability of the model (16).

## Granulation Criteria and Partial Order Relations

A **granulation criterion** is defined as the combination of (i) a **partial order relation** that specifies the ontological relations between granular parts across different levels of granularity and (ii) the specific **category of entity** to which this relation applies. A **partial order relation** (17) is a binary relation that is transitive, reflexive, and antisymmetric, such as the *class-subclass*, *parthood*, or *derived-from* relation. When granulation criteria are applied consistently, partitioning activities results in hierarchical partitions, i.e., **granular partitions** (18–21). Each particular granular partition can be represented as a **granularity tree**, with its nodes and leaves being granular parts (21–23), organised into different **levels of granularity**, with the root being the coarsest level, all its immediate children the next finer grained level, and so forth. All granular parts of a given level are thereby pairwise disjoint, but exhaustively represent the partitioned whole (21).

Depending on the chosen category of entity, the same partial order relation yields different types of granularity trees. The parthood relation, for instance, when applied to physical objects, yields material partonomies, when applied to documents and datasets information partonomies, and to activities and events process partonomies, and thus three different types of granularity trees, i.e., three **granular perspectives** (following Keet's general theory of granularity (24–26))

## Bona Fide vs. Fiat Boundaries and Causal Unity

Partitioning a referent system involves identifying **boundaries** that demarcate a foreground granular part from its background. Depending on the criteria applied for locating this boundary, the resulting granular parts either reflect the granular structure inherent in the referent system, and are demarcated by **bona fide boundaries**, or they only reflect mind-dependent criteria applied by the modeller, depending on fiat cognitive and contextual decisions of the modeller, and are demarcated by **fiat boundaries**. When consistently applied, these boundary criteria thus result in either **bona fide** or **fiat granularity trees** (16).

Distinguishing between bona fide and fiat boundaries is complex. The discussion has evolved from purely physical criteria (27–35) to more nuanced notions of **causal unity** (16,36–38). The latter defines how parts are unified into coherent wholes, each resulting in a specific family of granular perspectives:

1. **Causal unity via internal physical forces** maintains structural integrity against strengths of attractive or destructive forces via fundamental forces of strong and weak interaction, covalent bonds, ionic bonds, or metallic bonding. With its focus on physical forces, the resulting granular perspectives are associated with a **spatio-structural frame of reference** that focuses on inventorying **what is given** in a particular point in space and time by referring to the spatio-structural properties of the part (16). The following two subcategories can be distinguished that supervene on causal unity via internal physical (37,38):
   a. **Causal unity via physical covering** unifies the part against its surrounding, such as a plasma membrane of a cell.
   b. **Causal unity via engineered assembly of components** unifies the part through screws, glues, weld seams, and fasteners.
2. **Causal unity via bearing a specific function** unifies the part and its subparts through being the bearer of a function and all its sub-functions (37,38). The resulting granular perspectives are associated with a **functional frame of reference** that focuses on making **reliable predictions** of what can or necessarily will happen in the future by referring to the part's dispositions (16).
3. **Causal unity via common historical/evolutionary origin** unifies the part through a common historical or evolutionary origin that it shares with all of its subparts (e.g., developmental, genealogical, and evolutionary lineages) (37,38). The resulting granular perspectives are associated with a **historical or evolutionary frame of reference** that focuses on making **reliable retrodictions** of **what has happened** in the past by referring to the entity's (evolutionary) history (16).

These types of causal unities are critical to ontology design and semantic modelling, allowing for models that can be reasoned over, also by machines, across frames of reference.

# Multidimensional Granularity and Integrated Perspectives

Given the multidimensional granular complexity of real-world systems, any referent system can be partitioned in multiple valid ways, each based on different partial order relations and causal unity criteria. Proper modelling, then, involves explicit and **intentional selection** of which perspectives to represent.

Sound modelling often integrates **multiple granular perspectives**, some of which are **bona fide** (for ontological grounding), others are **fiat** (for usability or cognitive interoperability). Each serves a different epistemic, cognitive, or technical purpose.

# Applying Granularity Theory and Causal Unity Criteria to Structure Levels of Biological Organisation

With this foundation, we return to the metaphor of the levels of biological organisation and the idea of structurally and functionally stable building blocks. By applying logically and ontologically consistent granulation criteria and by focusing on the criterion of causal unity via physical covering, but at the same time considering the two functional and historical/evolutionary criteria, we can identify the following foundational types of **physical coverings** that demarcate **bona fide building blocks** relevant in the life sciences. The physical covering functions as a **barrier** that protects an inside milieu from the outside milieu, thereby establishing a micro-ecosystem inside the building block. In biology, the covering is itself a physical object that not only provides the surface boundary of the building block, but is also the bearer of the dispositions with which it **interacts** and **communicates** with its surrounding, functioning as the building block's **interface** to the world (16,39). The following types of physical coverings and their associated building blocks can be distinguished (16):

1. **Electron shell**: atoms, molecules

   *Portions of matter such as water in a glass do not form new types of building blocks but are aggregates of molecules, as a portion does not possess a single electron shell shared by all its molecules.*

2. **Plasma membrane**: single-membrane-enclosed entities (e.g., many organelles and all prokaryotic cells), membrane-within-membrane entities (i.e., eukaryotic cells)

   *Prokaryotic and eukaryotic cells can form aggregates (tissues), but a tissue does not represent a new type of building block, as it is an aggregate of cells that does not necessarily possess its own membrane.*

3. **Epithelium**[1]: epithelially delimited compartments (i.e., most organs), epithelially delimited multicellular organisms (i.e., organisms that possess an epidermis)

---

[1] a continuous single-cell layered sheet of tightly packed polarized cells, with a basal lamina at its basal surface that acts as a filter, comparable to a plasma membrane, but at the level of cell clusters

*Multicellular organisms can form aggregates (populations), but a population does not represent a new type of building block, as it is merely an aggregation of organisms that typically does not possess its own physical covering*

The resulting granular perspective is derived from a **granulation criterion** that applies the **physical parthood** relation as its underlying **partial order relation** to the different categories of **building blocks** introduced above, which combine all three criteria of causal unity. The granularity perspective comprises the granularity levels of *atoms > molecules > single membrane entities > membrane-within-membrane entities > epithelially delimited compartments > multicellular organisms with epidermis*. The hierarchy reflects not only a bona fide physical granular perspective, but also functional integration and evolutionary identity. It acts as a **foundational metamodel** in the life sciences, aligning and integrating multiple granular perspectives across domains (16).

For instance, a eukaryotic cell is a spatio-structural bona fide object (membrane-bound), a functional bona fide unit (self-regulating, self-maintaining), and as an instance of a cell type and as part of an evolutionary lineage that evolved to this cell type, an evolutionary bona fide unit.

This reveals how **evolution of new coverings** enabled emergence of new building blocks, driving the **evolution of complexity** through differentiation and combination of building blocks, much like Herbert Simon's *watchmaker* metaphor (40), in which stable subassemblies (= building blocks) accelerate complex system formation. This dynamic also echoes at the genetic and developmental level. For example, the *Character Identity Networks* (41,42), which stabilise development pathways and foster building block innovation.

## Granularity in Modelling Scientific Data, Information, and Knowledge

Regarding **modelling** of **data**, **information** and **knowledge**, and especially when creating FAIR **digital twins** (58) of a given real-world referent system, it becomes clear that the ontologically and logically sound representation of the referent system's internal granular structure is key to the scientific value as well as the machine actionability of the model. In other words, the world's granularity demands equally granular models, both at the level of empirical data and of theoretical and conceptual knowledge. All aspects of this internal granular structure that are relevant to the purpose and anticipated usage context of the model must be modelled as **bona fide granularity trees**, whereas (explicit and recognised) fiat granularity trees may be added to the model to support the model's usability, thereby taking into account the cognitive perspectives and limitations of humans and their socio-cultural requirements and constraints. The hierarchy of bona fide building blocks introduced above thereby provides the bona fide granularity perspective that ontologically and epistemically functions as the **backbone**, a **granular grid** or **connector**, to which modellers can refer to during modelling and analytics to interconnect and integrate various bona fide and fiat granularity trees and their corresponding frames of reference. By modelling FAIR data and knowledge along granular perspectives, we enable scalable reasoning, modular reuse, and integration across disciplines.

If scientific data and knowledge as models of real-world referent systems are necessarily granular themselves, with their explicit **information granularity** reflecting aspects of the granularity inherent in their referent systems, an obvious question is: **How does information granularity affect the FAIRness of** **a model? What are the unity criteria for demarcating *bona fide* granular parts of data and knowledge? Which are the building blocks of an information granularity hierarchy? What is the relation between FAIRness and granular data and knowledge?** To begin to answer these questions, we next introduce the semantic modularisation principle and the concept of semantic units.

# Semantic Modularisation and the Concept of Semantic Units

## The Semantic Modularisation Principle

*"Because human thoughts are combinatorial (simple parts combine) and recursive (parts can be embedded within parts), breathtaking expanses of knowledge can be explored with a finite inventory of mental tools."* —Steven Pinker ((43), p.360).

Drawing on the metaphor of **natural language and concept formation**, humans manage conceptual complexity by coining new terms that serve as semantic handles for complex parts (e.g., "*ecosystem*," "*quark*," "*prion,*" or "*peer review*"). The **principle of semantic modularisation** (7) brings this cognitive strategy into the domain of data and knowledge representation. It does so by partitioning complex FAIR datasets into **semantically meaningful, logic-aware, and independently addressable information units** that can be individually queried, annotated, reasoned over, or ignored, depending on the task at hand. This modular structure ensures:

- **Logical heterogeneity across subsets**: Each module can implement its own explicit logical framework, such as OWL (Description Logic), First-Order Logic, or even logic-free representations, enabling distributed reasoning across heterogeneous components.
- **Semantic coherence within subsets**: Each unit is self-contained, with its content being internally consistent and meaningful to a domain expert (as opposed to some triples that are not meaningful to a domain expert; cf. triple marked in red in Fig. 2).
- **Independent addressability**: Each module is uniquely identifiable and can be reused, cited, or queried independently.
- **Cognitive interoperability and contextual navigability**: Subsets are human-interpretable, navigable, and interlink into a coherent semantic structure, supporting contextual exploration.
- **Reusability and composability**: Modules can be recursively nested, forming complex structures while preserving modular semantics.

By decomposing datasets and information into such modules, semantic modularisation introduces **logical transparency and scope reasoning** at the granularity level of the unit. It explicitly defines epistemic boundaries, enabling hybrid reasoning practices that integrate formal logic with human interpretation. Each subset functions as a **FAIR and CLEAR information unit**, transforming data, information, and knowledge into cognitively interoperable and semantically rich artefacts (6).

Importantly, semantic modularisation is **technology-agnostic**. It defines a **conceptual strategy** rather than prescribing a specific technical implementation, be it RDF/OWL, property graphs, or SQL-based tabular data. This decoupling is critical, as technologies evolve, but the challenges of modelling and communicating semantic complexity remain constant. Semantic modularisation therefore offers a durable, infrastructure-neutral strategy for representing scientific meaning.

## The Semantic Unit Concept

Given that semantic modularisation provides the architectural principle for decomposing complex data and knowledge into modular, logic-aware units of information to deal with information complexity, then **semantic units** provide its structural implementation (7,8). Semantic units function as the **atomic and composite carriers of semantic content**, enabling information to be made **composable, traceable, and reusable** across different levels of abstraction and across technological ecosystems.

A **semantic unit**, from this perspective, is a semantically meaningful, structured, and independently identifiable subset of a data, information, or knowledge representation. It is typically implemented as a subgraph in a knowledge graph, a group of relational records, or a JSON object, but it is **not limited to any one formalism** (8).

Each semantic unit has a **globally unique and persistent identifier** (GUPRI) that enables referencing, reuse, citation, commenting on, and linkage of the unit's semantic content. This GUPRI can then be used to make statements about the semantic unit, including specifying all relevant metadata such as provenance, license, and a specification of the logical framework and data schema (i.e., schema metamodel) applied during modelling. Each semantic unit also instantiates a **semantic unit class**, which specifies the type of semantic content it contains, which can be used in search and browse activities, including **contextual exploration** (6) (Fig. 1). This makes semantic units **first-class FAIR digital objects** with **epistemic, computational, and archival significance** that meet the **CLEAR Principle** (6).

Semantic units are designed to support both machine-based operations (e.g., querying, reasoning, indexing) and human-centric tasks (e.g., browsing, annotation, interpretation). Two top-level categories can be distinguished: statement units and compound units.

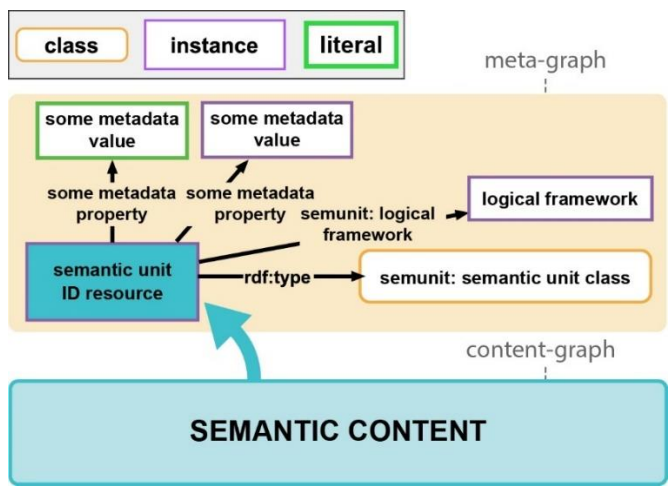

**Figure 1: Core architecture of a semantic unit, exemplified within an OWL/RDF framework.** The semantic content of a semantic unit is always represented by a GUPRI, i.e., the semantic unit resource. Depending on the technological implementation, this GUPRI somehow identifies the semantic content. In an OWL/RDF framework, this could be a Named Graph, with the Named Graph GUPRI being also the semantic unit's GUPRI. This GUPRI instantiates a corresponding semantic unit class. Metadata, including provenance, license specification, and the specification of the logical framework that has been applied, if any, for the modelling of the semantic content, are specified within the meta-graph, whereas the semantic content itself is located in the content-graph Named Graph. Every class and instance also possesses its own GUPRI.

Semantic units are designed to support both machine-based operations (e.g., querying, reasoning, indexing) and human-centric tasks (e.g., browsing, annotation, interpretation). Two top-level categories can be distinguished: statement units and compound units.

## Statement Unit

Atomic data elements, such as individual database cells, RDF resources, or key-value-pairs, are not semantically meaningful in isolation. Much like words in natural language, they derive meaning only through their syntactic structuring and contextual integration (12). To enable modular, reusable, and interpretable modelling, **statement units** treat **individual propositions**, i.e., minimal, complete, and independent units of meaning, as the **smallest reusable building blocks** in FAIR and CLEAR data modelling.

When applied to a dataset, **semantic modularisation partitions it mathematically** into disjoint statement units, such that each triple, cell, or data field belongs to exactly one unit (7). Each **statement unit** captures a singular, cognitively coherent proposition about a referent system, its **semantic content**, forming a **unit of information that is also meaningful to a human user** (8) (Fig. 2C-F). A statement unit may contain a subject and one or more objects.

The semantic content expression is located in the unit's content-graph and represents the unit's actual meaning, organised according to a predefined data schema, which reflects a corresponding statement unit class. This **semantic content** can be realised in **multiple structural forms and technical implementations**, such as a reasoning-capable OWL/RDF-based subgraph, SQL row grouping in a relational table, JSON object or structured document, RDF-based subgraph or SQL row grouping that applies the Rosetta Statements approach of modelling natural language statements instead of real-world referent systems (44), or even unstructured natural language sentences as vectorised or vector-embedded text snippets (Fig. 2).

What unifies these forms is not their syntax or technology, but their **semantic equivalence**. Each form is a token model, representing a specific proposition derived from a real-world referent system. These representations are **transitive to an underlying natural language token model** (e.g., "*Parasite X has a mass of 24.76 grams*" (Fig. 2A)) and aligned with a **natural language metamodel** that generalises the roles and relationships involved (e.g., "*OBJECT has a QUALITY of VALUE UNIT*" (Fig. 2B)). The natural language metamodel can also be used for **normalising** and thus **harmonising** all natural language token models modelling the same type of data or knowledge (e.g., "*The mass of Parasite X is 24.76 grams*" is normalised to "*Parasite X has a mass of 24.76 grams*").

From a modelling perspective, a statement unit's semantic content, either modelled as a tabular structure (Fig. 2C), a natural language sentence (Fig. 2D), an OWL-based graph (Fig. 2E), or a Rosetta Statement (Fig. 2F), can be compared with its corresponding natural language statement (Fig. 2A). Both are token models that model the same set of entities, properties, and relationships of the same referent system. They only differ in type (a scientific claim as a nanopublication, a machine-actionable instruction for an algorithm container, etc.), the use of the specification language, and the format. Consequently, the entities modelled can be aligned across these token models, and the models can be transferred into one another (see *schema crosswalks* in (12)).

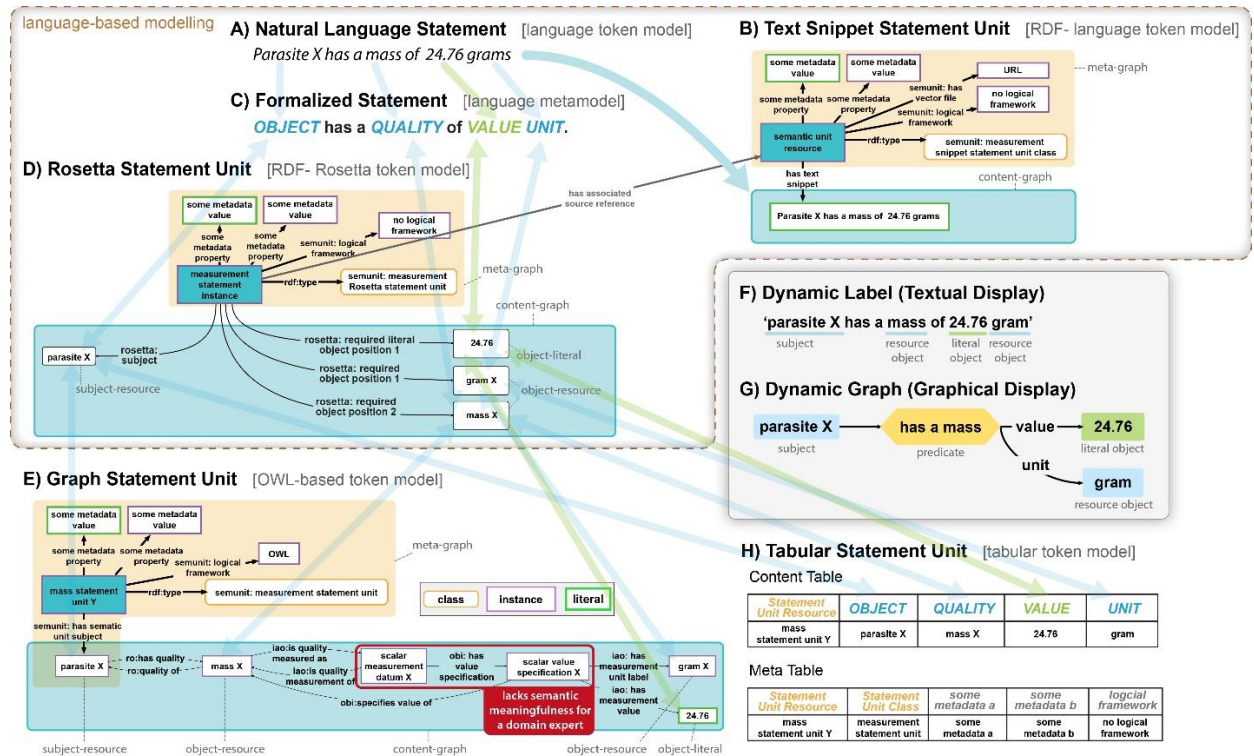

**Figure 2: Semantic transitivity between statement units carrying semantically equivalent content. A)** A human-readable token model in the form of an assertional statement, modelling the mass measurement of a particular parasite X. **B)** The statement from A), modelled as an RDF-based token model in the form of a text snippet statement unit. The unit's content-graph, denoted within the blue box, carries the unit's semantic content (cf. Fig. 1), modelled as a natural language string. In text snippet statement units, the string is linked to the unit's resource as the value of a data property. In other types of statement units, such as Rosetta Statement Units (D), the content is modelled as a subgraph in its own Named Graph. In either way, the content is connected to the semantic unit resource (bordered blue box) that is a GUPRI representing the content in the graph. The text snippet is vectorised, and the vector file is linked to the unit's resource along with all embedding-related metadata. Additional metadata include the reference to the source text, if applicable, along with start- and end-offset information to track the original location of the text in the publication. One can argue that the text snippet statement unit GUPRI is an identifier for its vector. The peach-coloured box encompasses the unit's meta-graph, which explicitly denotes the unit's resource (bordered blue box) as an instance of the '*measurement snippet statement unit*' class, therewith specifying the type of information the unit carries. The meta-graph also contains various metadata triples, here only indicated by *some metadata property* and *some metadata value* as their placeholders for detailed provenance, time stamps, etc. relating to this semantic unit. It also indicates which logical framework, if any, has been applied in the modelling process and thus, whether and which reasoning framework can be applied to the unit's semantic content (here: no logical framework has been applied, since the content is modelled in natural language). **C)** The language metamodel that is instantiated by statement A), in the form of a formalised statement, in which the subject and object positions have been classified and generalised (e.g., '*parasite X*' to '*OBJECT*' and '*mass*' to '*QUALITY*'). **D)** The statement from A), modelled as an RDF-based statement unit, with the semantic content modelled as a Rosetta statement, attempting to model the grammatical and syntactical structure of the corresponding natural language metamodel C) (for Rosetta Statements, see (44)). Models A)-D) represent language-based models in the sense that they either use natural language to represent the unit's semantic content (A-C) or use RDF/OWL to model a natural language statement (D). **E)** The statement from A), modelled as an OWL-based token model in the form of a statement unit. The semantic content is modelled as an OWL graph, with '*parasite X*' as its subject and '*mass X*', '*gram X*', and the numerical value '*24.76*' as its objects. The meta-graph identifies '*parasite X*' as the unit's subject resource. Since the semantic content is modelled in OWL, OWL is denoted as the unit's logical framework. Highlighted in red within the content-graph is an example of a triple that is required for reasoning but that lacks semantic meaningfulness for a domain expert. Contrary to D), this OWL-based model is a digital twin of its real-world referent—it attempts to model the relationships between real-world entities rather than modelling a natural language statement. D) and E) can be linked to a set of Text Snippet Statement units, with each Text Snippet Unit functioning as a source reference. This way, a given proposition that is modelled as D) or E) can have more than one source reference specified in the knowledge graph. **F)** The dynamic label associated with the measurement statement unit class. **G)** The dynamic graph associated with the measurement statement unit class. **H)** The statement from A), modelled as a tabular token model for use in, for instance, a MySQL database. The blue and green arrows indicate the alignment of resource and literal slots across the different models, indicating their semantic equivalence, with the Rosetta Statement schema functioning as a reference schema. Every class and instance also possesses its own GUPRI.

The language metamodel thereby acts as the semantic blueprint from which schema metamodels can be derived for each technical realisation: a **SHACL shape** (45) for OWL/RDF-based knowledge graphs, a **SQL Data Definition Language (SQL DDL) schema** (46,47) for relational databases, or a **JSON Schema** for JSON-based systems. This ensures that **semantic equivalence** can be preserved across different representations, and that both **machine-actionability** and **cognitive interoperability** are maintained. While natural language based statement units and their corresponding machine-actionable variants differ in format, they remain **semantically equivalent** and can be **translated or rendered interchangeably** via schema crosswalks (12).

To support **cognitive interoperability**, each statement unit should also define one or more **display patterns**, such as: **Dynamic labels** for human-readable textual representations in, for example, HTML views (see Fig. 2G) or **dynamic graphs** for their graphical display (Fig. 2H). This ensures that human users, whether domain experts, curators, or analysts, can interpret the unit's semantic content without needing to understand its underlying technical representation.

For a taxonomy of statement units, ranging from assertional and contingent to prototypical, universal, epistemic, conditional, and logical argument statements, see (7).

## Compound Unit

While statement units correspond to the first metaphor, as they coin GUPRIs comparable to defining new terms in linguistics to identify semantically meaningful parts within a given dataset, compound units realise the second: modelling hierarchical levels of organisation. They allow for the **granular nesting of semantic content**, forming the scaffolding that mirrors the granular structure of the corresponding referent system. Whereas statement units form the level containing the semantic content, compound units organise statement units into **semantically meaningful collections**, forming coarser-level organisational units. Like every semantic unit, a compound unit possesses its own GUPRI that instantiates a corresponding class (7,8) (taking up the first metaphor again: new terms are defined by already existing terms). However, compound units do not carry semantic content themselves—they do not possess a content-graph. Instead, they list in their meta-graph all the statement units and compound units that are part of their collection by referencing their GUPRIs (Fig. 3). In other words, the semantic content in terms of data and knowledge is restricted to the layer of statement units, and compound units only reference it through the GUPRIs of statement units but do not carry semantic content themselves.

Compound units allow modular organisation of a dataset into semantically meaningful and logically cohesive structures, with compound units within compound units, thereby reflecting the **granular nature of the referent systems** that they model.

Different types of compound units can be distinguished (7,8): **Item units** group statement units that share a common subject. **Item group units** group item units that are linked via statement units, where one item unit shares its subject with the subject of the linking statement, and one of this statement's objects is the subject of another item unit. **Granularity tree units** group all statement units that model a shared partial order relation of the same referent system, using a common granulation criterion. A **granular item group unit** is a granularity tree unit that includes for each node in its tree the corresponding item unit. A **context unit** groups statement and compound units of the same referent system by frame of reference. For example, information about a publication organised into three distinct context units: bibliographic metadata about the publication (context: publication as an item), procedural statements describing research activities (context: research process), scientific findings as result of the research activity (context: data and knowledge about the research subject). A **logical argument unit** relates premises and conclusions across multiple statement units. **Standard information units** are collections of semantic units that conform to a community standard for a particular domain, such as product sheets in industry that are based on specific DIN standards. For more compound unit types, see (7,8).

Compound units thus provide the conceptual structure for modelling and representing information granularity, enabling FAIR and CLEAR representation across nested, multi-scale systems.

## A) Graph Compound Unit

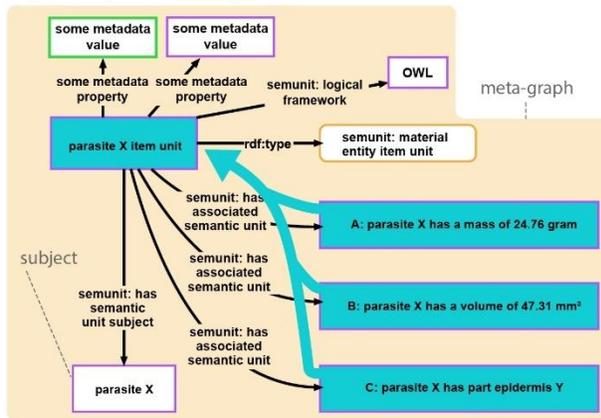

## B) Tabular Compound Unit

| Compound Unit Class | Compound Unit Resource | subject | associated semantic unit | associated semantic unit | associated semantic unit | some metadata a | some metadata b |
|---|---|---|---|---|---|---|---|
| material entity item unit | parasite X item unit | parasite X | A | B | C | some metadata a | some metadata b |

**Figure 3: Compound unit example. A) A graph-based representation of the compound unit**. The compound unit resource is denoted as 'parasite X item unit' and, as a compound unit of the type *item unit*, encompasses all statement units that share 'parasite X' as their subject, providing a description of 'parasite X'. Compound units do not possess their own content-graph and thus do not directly carry their semantic content. Their semantic content can be obtained by merging the content-graphs of their associated statement units (here indicated by the blue arrow). The compound unit is represented in the knowledge graph via its meta-graph (depicted in the peach-coloured box), which links the resources (here: A, B, and C, shown with their dynamic label for readability of the figure) of the associated semantic units to its compound unit resource. **B) A tabular representation of the same compound unit**.

# Semantic Units across Information and Technical Layers

Semantic units, as introduced above, function as first-class carriers of semantic content. Beyond their structural and computational role, however, they serve as **conceptual bridges** between **human-centric representations** and **machine-actionable semantics**. This dual function requires a clear distinction between two intersecting dimensions:

1. **The information dimension: Semantic units as machine-interpretable and cognitively interoperable models**

   At their core, semantic units model information, not technical structures. They **partition data and knowledge** into units of information that are not only machine-interpretable but also **semantically meaningful to human users**, with statement units serving as the finest grained building blocks of **FAIR semantics**. Each semantic unit is classified into a specific type of information unit, associated with its corresponding data schema, which functions as its metamodel. This classification ensures propositional interoperability across all its instances (12). From this perspective, the technical implementation of a semantic unit is irrelevant. What matters is its capacity to carry semantically coherent content, with a focus on optimising **computer-information and human-information interaction**.

2. **The technical implementation dimension: Transposing semantic units across technologies**

   Given a well-defined language metamodel and semantically equivalent schema metamodels, we can translate semantic units into various technical formats, each corresponding to a specific data schema. These may include SHACL shapes for RDF/OWL graphs and SQL DDL schemata for relational databases (see Fig. 1C,D). Each representation is a **manifestation** of the same semantic content, analogous to the Functional Requirements for Bibliographic Records (FRBR) model of a *work* (abstract content), its *expression* (intellectual realisation), and its *manifestation* (concrete format) (48). Just as *Romeo and Juliet* can appear as a hardcover or eBook, a particular statement unit may be concretised as an OWL/RDF subgraph in a knowledge graph or a row in a table in a relational database. Ideally, these different **technical concretisations** are linked via a schema service or registry, where language metamodels are defined and versioned, crosswalks between different statement unit manifestations and their corresponding language metamodels are maintained, and clients and tools can negotiate optimal formats for specific use cases, all supporting their propositional interoperability (12). Projects like Abstract Wikipedia (49) exemplify this model by aiming for enabling formalised propositions to be rendered into multiple languages and formats, grounded in structured semantic models. In this technical dimension, the emphasis shifts to machine-usability and interoperability with technical infrastructures while still being transitive to their corresponding natural language models, supporting optimal **computer-computer and human-computer interaction**.

By distinguishing between the information and the technical implementation dimension, we position **semantic units as conceptual variants** across formats and infrastructures of an underlying **invariant natural language metamodel**. This dual role is central to enabling technology-agnostic modularisation of data and knowledge, cognitive interoperability combined with machine-actionability, and sustainable FAIR and CLEAR data infrastructures. This conceptual architecture, grounded in abstraction, granularity, and modular reasoning, prepares us for the next step: implementing semantic units as FDOs.

# Semantic Units as Semantic FAIR Digital Objects and Information Granularity

## Semantic Units as Semantic FAIR Digital Objects

Semantic units, as introduced above, are granular, independently identifiable, and cognitively meaningful modules of scientific content. Here, we explore how these can be serialised as **semantic FAIR Digital Objects** (semantic FDOs), and what implications this has for their role within a future Internet of FAIR Data and Services (IFDS).

FDOs aim to make digital objects **findable, accessible, interoperable, and reusable** in a persistent and machine-actionable way. By embedding semantic units into this architecture, we now also bring **granularity**, **logical transparency**, and **cognitive interoperability** to the forefront of FAIR design, thereby bridging human understanding and machine-actionable semantics. Based on this **semantic unit framework**, the envisioned semantic FDO architecture aligns both with the FAIR Principles and the CLEAR vision of cognitively interoperable and contextually explorable data and knowledge, and this not just at the level of datasets, but at the level of **information granularity** itself.

### Mapping Semantic Unit Types to semantic FDO Types

We propose a direct mapping between **semantic unit types** and **semantic FDO types**, extending and revising our previous work on FAIRness and CLEARness of semantic content (6,8):

- **Statement Units → Statement FDOs**
  Statement FDOs are minimal, self-contained units of propositional content. They may assert facts, rules, hypotheses, or instructions. Each statement FDO contains a formalised statement along with sufficient metadata to support both human comprehension and machine usability. Statement FDOs can be realised for instance as Research Object Crates (RO-Crates), i.e., **Statement RO-Crates**, when the corresponding statement unit is based on a **tabular format**, or as a **Nanopublication** (8,50–52) when it is based on an **OWL/RDF graph**, and other technical serialisations are also possible.
- **Compound Units → Nested FDOs**
  Nested FDOs encapsulate semantically meaningful collections of multiple statement units (or other compound units), such as logical arguments, procedural clusters, or context-specific packages. Analog to statement units, compound units can be realised as RO-Crates, i.e., **Nested RO-Crates**, or as extended Nanopublications, i.e., **Nested Nanopublications**.

This mapping reflects the **compositional nature** of data and scientific knowledge and enables semantic FDOs to represent content at multiple levels of information granularity.

## Format Flexibility: RO-Crates and Nanopublications

The implementation of semantic units as FDOs can vary across technological ecosystems, depending on format preferences, tooling, and community practices. After all, data and knowledge representations are models, and each model serves a specific purpose and anticipated **usage context**, including readily usability with widely used analysis tools and compliance with established formats.

- **Nanopublications** typically serve as a serialisation method for **statement FDOs** in the OWL/RDF ecosystem. A Nanopublication wraps a single claim (assertion) with its associated provenance, context, and publication metadata in a structured, machine-readable format. Since **their structure allows referencing the GUPRIs of several statement FDOs** in their *head* Named Graph while leaving their *assertion* Named Graph for compound FDOs empty (8), Nanopublications can be extended to also represent OWL/RDF-based **compound units**. To avoid confusion due to the *nano* prefix, these could be subclassed or labeled explicitly as **Nested Nanopublications** (e.g., using a Boolean '*hasCompoundStructure*': *true*).
- **RO-Crates** serve as a serialisation method for **statement** and **nested FDOs** in **tabular or document-based ecosystems**. They allow for semantically rich packaging of datasets, metadata, workflows, and software, linked via persistent identifiers and described using JSON-LD. While usually used for packaging entire datasets, **RO-Crates can also be used for serialising individual tabular statement units**. In such cases, a crate would include a file (e.g., a CSV row or cell cluster), a data schema (e.g., SQL DDL or schema.org), metadata identifying the statement unit type, and a human-readable rendering (e.g., label or sentence).

This format flexibility reinforces the technology-neutral design of the semantic units framework and enables **cross-ecosystem semantic parity** between semantic FDO implementations. The use of Nanopublications and RO-Crates to document semantic units is not exclusive, nor do they define the semantic FDO concept. Instead, they represent **technology-specific manifestations** of a broader architectural pattern: the FAIR-aligned, machine-actionable modularisation of digital knowledge.

## FAIRness, Modularity, and Granularity

Currently suggested FDO implementations run the risk of suffering from a **lack of granularity**, treating large datasets or metadata bundles as monolithic units. In contrast, semantic units offer a **granular structuring mechanism** that aligns FAIRness with semantic tractability, allowing FAIRness to be assessed, enforced, and reused at the level of individual statements or semantically meaningful collections of statements.

Each semantic unit, when serialised as a semantic FDO, should clearly specify **what kind of information** it carries (e.g., assertional statement, universal statement, compound item, compound granularity tree, etc.). Each semantic FDO should also specify **how it is structured** (data schema) and **which logical framework**, if any, underlies its formal semantics. Additionally, it should indicate **what level of granularity** it occupies within a larger granular information architecture.

This makes FAIRness of data and knowledge not just a metadata issue, but a **semantic property** that is linked to its cognitive interoperability, composability, and contextual transparency.

## Metadata Requirements for Semantic FDOs

One of the fundamental challenges in today's digital research ecosystem is the **isolation of technological domains**. We operate within fragmented islands: RDF/OWL-based knowledge graphs, Python-based scientific workflows, relational databases, and spreadsheet systems. Each is optimised for specific tasks, communities, or historical legacies.

**Semantic units**, when serialised as semantic FDOs, can serve as **bridges** between these islands. By enforcing persistent identifiers (GUPRIs), explicit schema declarations (via schema metamodels), and crosswalks between schemata and formats, they enable statements and datasets to **move between ecosystems** without semantic loss. This decouples **semantic content** from any one implementation stack, supporting the **portability of scientific knowledge** across infrastructures.

To fully support both FAIR and CLEAR principles, each semantic FDO should carry structured metadata that includes (but is not limited to):

1. **Statement FDO**
   - **Authorship and Provenance**
     The **creator** of the digital object, and the **author** of its content (if different), as well as temporal and institutional provenance. This enables proper attribution and assessment of trust.
   - **Schema Declaration**
     A resolvable identifier of the schema used (e.g., SHACL, SQL DDL, JSON Schema). This enables machine-actionable validation, composability, integration, and interpretation.
   - **Logical Framework**
     Specification of the formal logic employed (e.g., OWL DL, First-Order Logic, or logic-free). This is essential for informing reasoning scope and inference compatibility.
   - **Statement Typology** (if a statement unit)
     Clearly identify the **semantic type** of the proposition: Assertional (*This swan is white*), contingent (*Swans can be white*), prototypical (*Swans are typically white*), universal (*All swans are white*), interrogative (*Is this swan white?*), directive (*Make this swan white!*) and directive conditional (*If you see a black swan, notify Anna!*), interrogative (*Are there any black swans?*), statements about epistemic believes (*Anna believes: All swans are white*), etc.
   - **Human-readable Rendering**
     The ability to render statement unit's semantic content represented as a natural language (e.g., dynamic label, HTML representation). This ensures **cognitive interoperability**, allowing humans to interpret the semantic FDO's meaning without deep technical parsing, and supports user interfaces.

2. **Nested FDO**
   ○ **Inventory of the nested FDOs**
     A list of all types (i.e., classes) of FDOs that are part of the collection of the nested FDO, including indirectly associated FDOs, when other nested FDOs are part of the collection, providing an overview of the granular richness, information composition, and depth of the nested FDO.

## Toward an Internet of FAIR Data and Services

The long-term vision of an **IFDS** relies on a **network of modular, interoperable, machine-actionable entities** that can be composed, reasoned over, and used across infrastructures. Semantic FDOs directly support this vision by enabling:

- **Granular and decentralised composability**: Nested FDOs and thus semantic FDOs referencing other semantic FDOs via GUPRIs enable partial reuse.
- **Interoperable reasoning**: Defined logical frameworks and schema validation.
- **Semantic transparency**: enabling alignment across domains and infrastructures.
- **Cross-ecosystem operability**: enabling FAIR workflows across RDF, tabular, and code-based systems.
- **Discoverable content**: Search and reuse based on semantic unit class and schema metadata, in addition to other search strategies (e.g., search via vector embeddings for text snippets, structured query via SQL or semantic search via SPARQL).
- **Cognitive interoperability and machine-actionability**: Human-friendly labels and interfaces, integrated at every level of granularity, satisfy human-actionability needs, and the grounding in schema metamodels machine-actionability needs.

These capabilities position semantic FDOs as **foundational building blocks for the next generation of open, machine-actionable (scientific) research communication infrastructures**, supporting not only data access and reuse, but intelligent interoperation and semantically coherent automation. The capacities enable machines to participate in scientific knowledge processes without compromising human interpretability.

# FAIR Digital Objects and Information Granularity: Learning from Biology

A central metaphor guiding this work is the analogy between the **hierarchy of levels of organisation within biological entities** and the **hierarchy of levels of organisation within information content entities**. Both hierarchies can be conceived to be organised around **granular, layered structures**, i.e., building blocks, who are demarcated in such a way that they are parts within several different granularity perspectives, providing a **mediating organisational framework for integrating all major bona fide granularity perspectives** that apply to their respective referent systems.

In biology, these building blocks range from atoms and molecules to cells, and multicellular organisms with an epidermis. Their demarcation criteria, while having a focus on physical covering, combine all three **causal unity criteria**, with their identity as spatio-structurally distinct entities is defined through **physical coverings** (e.g., electron shells, membranes, or epithelia) that demarcate internal from external, enabling local autonomy while permitting structured interaction. These boundaries are not arbitrary but reflect **bona fide physical demarcations** that preserve the entity's function, internal organisation, and causal coherence.

In the FAIR information ecosystem, a parallel structure emerges through **semantic units**, which are demarcated by **semantic unity criteria**, resulting in ***bona fide* semantic boundaries**. We define a *bona fide* semantic boundary in the context of information as the boundary that demarcates a semantically meaningful granular part of information (data, information, or knowledge) from its information background. Like with biological building blocks (e.g., electron shells, membranes, or epithelia), we can distinguish different categories of such *bona fide* semantic boundaries. The **smallest units of information are the terms, symbols, or phrases that represent individual entities** (i.e., objects, qualities, processes, dispositions, etc.) and **types of entities** (i.e., classes and other forms of groupings thereof). These are used in modelling when referring to the entities and types of entities of a referent system. While representing *bona fide* units of information, individual terms or resources referring to individual entities are *per se* not semantically meaningful to a human user, and consequently represent *fiat* granular parts of *bona fide* units of meaning. The smallest semantically meaningful units of information are statements (i.e., propositions).

The semantic boundaries are **analogous to the physical coverings** such as membranes and epithelia: they define **what kind of semantic content** they encapsulate and, via the specification of their underlying logical framework and data schema, **what kind of interactions in terms of reasoning and propositional interoperability** they support.

Importantly, **semantic FDO types** align with semantic unit types. Unlike in biology, however, where the physical structure is intrinsic, in FAIR systems the **technical wrapper (e.g., RO-Crate, Nanopublication) is independent** of the **semantic content**. A universal statement, for example, can be expressed as a text snippet, an RDF/OWL based Rosetta Statement, a reasoning-capable OWL-graph, a row in a table, or a JSON object (cf. Fig. 2B,D-F), serialised as either RDF-based Statement Nanopublication or Statement RO-Crate. Its **semantic identity**, not its technical implementation or serialisation format, determines its position in the hierarchy of information granularity.

Analogous to biological systems, the question is now what distinguishes an information building block from the other types of bona fide information units. **Can we identify semantic FDO types that take a central position within the information granularity and integrate various information granularity perspectives?**

We think that statement units, item units, and item group units satisfy this prerequisite, as they form three different and clearly distinct **levels of representational granularity**, with each information building block type being involved in various different information granularity perspectives (6,8). Each **level in the resulting information granularity hierarchy** corresponds to either (i) a symbol, identifier, or word representing individual entities or types of entities or (ii) a type of semantic unity:

- **GUPRI or textual representation**: An identifier, a literal associated with a defined datatype, or a natural language term, representing an individual entity or type of entity of a referent system.

- **Statement Units**: Carry propositional content (e.g., assertional, contingent, prototypical, universal statements).
- **Item Units**: Aggregate statements about a single subject.
- **Item Group Units**: Aggregate related item units based on semantic connections (i.e., statement units that connect item units).

The **semantic interface** (i.e., the corresponding entity/type or semantic unit type specification) of each information building block type informs both the human reader (via schema, labelling, and display patterns) and the machine (via schema metamodels, logical framework specification, and addressability). Just as **cell membranes** define the boundaries and functions of biological cells, entity and type specification as they are provided by **controlled vocabularies and ontologies** and by **semantic unit type specifications of information building blocks** define what can be known, accessed, or done with that information unit. In a complementary way, they can also be understood as analogous to an **application programming interface (API)** that provides a structured, transparent communication surface through which both humans and machines can interact with, query, and integrate information across systems.

## Comparison of Biological Levels Hierarchy and FAIR Information Levels Hierarchy

We can now align the **hierarchies of biological and our proposed stack of informational building blocks**, both of which follow a **cumulative-constitutive logic**, with building blocks from coarser levels being composed of building blocks from finer levels. Consequently, each coarser level is **ontologically dependent** on the building blocks of the directly related finer level. Figure 4 visualises the alignment between biological and FAIR information granularity hierarchies, illustrating how (ontology) term information building blocks and semantic-units-based information building blocks mirror biological building blocks in layered composition and boundary definition.

The **demarcation criterion** for the biological building blocks is **causal unity by different types of physical covering**, whereas for the information building blocks, it is **semantic unity by different types of semantic coherence**. At the finest level, we can align the levels of atoms and molecules with the level of (i) GUPRIs or natural language terms that comprise resources or strings referring to individual objects and properties and (ii) literals defined by datatypes. Both form the most fine-coarse units within their hierarchy.

Within the biological hierarchy, coarser units are distinguished from the finest level building blocks by the emerging property of **life**. In the information hierarchy, it is the emerging property of **semantic meaning**, and 'actionability' for a particular purpose.

The building blocks in both hierarchies are distinguishable by their **unity criteria**, e.g., prokaryotic cell by single plasma membrane and statement units by single proposition as semantic content. Each building block is **internally organised** and functionally integrated. It can be **identified**, **reused**, and **composed** into coarser units, and is part of multiple granular perspectives, effectively *'bringing information to life'*.

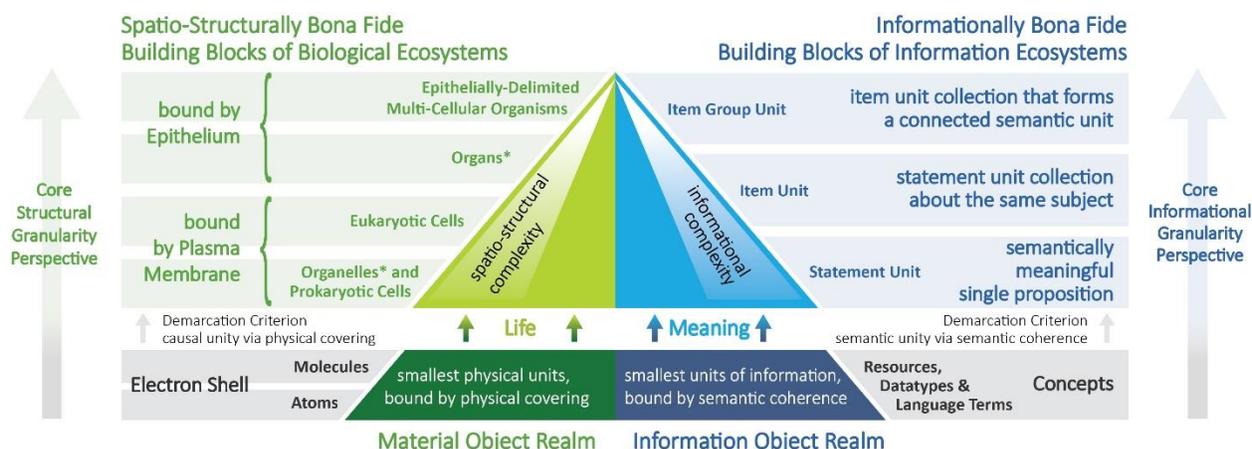

**Figure 4: Comparison of the Biological Levels Hierarchy (left) and the Information Levels Hierarchy (right).** Both hierarchies are based on corresponding building blocks that are demarcated by causal unity via physical covering (biological hierarchy: electron shell, plasma membrane, epithelium) or semantic unity via semantic coherence (information hierarchy: concept, semantically meaningful single proposition, statement unit collection about the same subject, collection of item units connected via statement units).

The aspect of integration with multiple granular perspectives is an essential property of building blocks, as formally defined by their capacity to integrate multiple granular perspectives while maintaining semantic coherence. As discussed earlier, the same referent system can be partitioned and modelled according to **multiple granularity perspectives**: A phenotype description may be organised into various statement units, which are organised into different item units that are grouped into a connected item group unit. A given statement unit can be part of one of the item units of the description and at the same time of one of the context units and one of the granularity tree units. Similarly, one of the item units can simultaneously be part of an item group unit and a granular item group unit. Each of these perspectives has its own **internal granularity**, but they often **interlock via shared building block units**. Thus, **information building blocks based on concepts** (referenced either via GUPRIs or natural language terms) **and on specific types of semantic units become the structural anchors** across which different information granularity perspectives can be integrated, navigated, and reasoned over.

The analogy between biological and FAIR ecosystems, where information building blocks are defined via semantic unity, suggests that terminology and schema registries should evolve to classify not only vocabularies, data formats, and schemata, but also the **semantic granularity level and type of information unit** they support. This would enable more modular, resilient, and evolving interoperable tooling, facilitating schema crosswalks, semantic validation, and composition across heterogeneous datasets.

# FAIR Semantics and the Granularity of FAIRness

## From Monolithic to Granular FAIRness

The FAIR Principles (2) are often interpreted and applied as if FAIRness is a binary or monolithic property, describing a dataset's global, binary status of being either FAIR or not. Yet, as both the complexity of datasets and our expectations for reuse grow, it becomes clear that this view is limiting. **FAIRness is neither Boolean nor flat**, but rather a **granular spectrum** (12). FAIRness can and should be evaluated **gradually** (to what extent an object fulfils each of the FAIR principles), **granularly** (at which level of information granularity and with which granular richness and depth), and from different perspectives (**perspectively)** (within which usage context or modelling frame of reference).

This chapter discusses the need for a **modular, multi-level model of FAIRness**, rooted in the granular architecture of semantic units. It argues that FAIRness should be assessed at the level of **individual statement units** (i.e., statement FDOs), **compound units** (i.e., nested FDOs), as well as at the level of entire datasets, each possessing varying degrees of granular complexity, modular reuse, and user relevance.

## Assessing FAIRness Across Granularity Levels

FAIRness is not a goal in itself. As stipulated in the original article, the principles serve the '*the ability of machines to automatically find and use the data, in addition to supporting its reuse by individuals'* (2). Any evaluation of 'FAIRness' should thus include the inherent question: '*will a machine know what I mean*'? Information exists at multiple levels of granularity. Consider two datasets, *A* and *B*, describing the same biological specimen. Dataset *A* describes only the organism as a whole, not including descriptions of its finer-grained anatomical parts. Independent of its lack of granular depth, each semantic unit in the description has a maximum FAIRness score of *X*. Dataset *B* includes both the organism and its organs, each described individually. Each semantic unit in these descriptions has a maximum FAIRness score of *X*. While their individual components are equally FAIR, dataset *B* contains a richer granular structure than dataset *A*. Should this granular complexity be reflected in its overall FAIRness score?

This example exposes a key challenge: how to evaluate and compare datasets that differ in both granular structure and scope. We propose a dual-scoring approach:

> *granular FAIRness = mean FAIRness* x *granular complexity*.

This allows modular content to be rewarded for granular depth and richness, without penalising simpler but well-curated datasets. Moreover, instead of providing a single FAIRness score for a given dataset, each type of semantic FDO that it contains, independent of being for instance a statement FDO or any type of nested FDO, can and should be independently evaluated. This modularity enables flexible scoring and use-specific filtering. For example, a statement FDO representing a universal claim might carry its own FAIRness score, which is inherited and aggregated by a nested FDO containing many such statement FDOs. Aggregation functions (mean, weighted, normalised) must be transparently reported.

Such evaluations are more than technical, as they reflect the **semantic structure** of the modelled domain and support multi-stakeholder needs. However, one can also argue that a single (granular) FAIRness score cannot do justice to the fundamental differences in FAIRness across semantic FDOs covering different levels of information granularity. As an alternative to a single quantitative FAIRness score or listing the FAIRness scores of all statement FDOs belonging to a nested FDO, a qualitative assessment could be provided. This could include the specification of all FDO types that are directly or indirectly part of the collection of FDOs of a given nested FDO. A query traversing the paths of associated FDOs could collect the respective FDO class affiliations, which could be listed with the nested FDO. This information would provide a good overview of the granular depth and granular richness that this nested FDO covers.

Moreover, if each FDO class specifies its corresponding schema specification (using a GUPRI for the schema) and an operations/functions service (see discussion below) lists all available operations/functions that are interoperable with a given schema, we also get a good account of the FDO's machine-actionability and thus AI-readiness.

## Interoperability and Content-Aware FAIRness

A common misconception in FAIR evaluations is that interoperability can be guaranteed by simply aligning metadata. However, **true semantic interoperability requires additional content-level alignment** (12).

Consider a dataset where all metadata rows share the same ontology annotations and underlying metadata schema specification. This ensures metadata interoperability. But unless the data specifies its underlying data schema, reflects meaningful semantic types, specifies its logical modelling framework, and the data instances align with **cognitively accessible content structures**, the dataset itself is not truly human- and machine-interoperable, and thus reusable in practice. The dataset then **may be FAIR regarding its metadata but is neither FAIR nor CLEAR regarding its data** (6). Templates and schema services can help, but their quality, granularity, and transparency vary significantly.

Therefore, FAIRness assessments should include checks for **semantic unit type** (statement, item, item group, etc.), **schema metamodel validation**, and **alignment with natural language equivalents**, ensuring cognitive and semantic interoperability. Obviously, FAIR metadata have an added value in themselves, even if they explicitly state that the data they refer to are not FAIR (yet), as it can lead to interaction and potentially in FAIRification of (parts of) the data set itself.

## Stakeholder- and Perspective-Sensitive FAIRness

Not all users care about FAIRness at the same level. Different users assess FAIRness from different perspectives, have different FAIRness expectations based on different **information usage contexts**. If a user plans to conduct a very specific analysis on the data using a specific software tool for running the analysis, they want the data to be readily usable with *that tool or that particular visiting algorithm*.

**Data stewards** prioritize schema transparency and provenance, **researchers** prioritize citation, modular reuse, and traceability, **machines** require logical consistency and, ideally, formal semantics with well-defined inference boundaries, and **archivists** prioritize long-term semantic coherence across evolving infrastructures.

FAIRness is not only structural but epistemic. A given semantic unit may score highly in one perspective (e.g., for general findability) but lack specificity in another (e.g., fine-grained reusability). Recognising these differences enables not only targeted improvement recommendations, but also FAIRness dashboards that map semantic coverage, and use-case driven validation templates.

A granular-aware FAIRness model allows FAIRness to be (i) **context-sensitive**, allowing stakeholders to target evaluations at relevant levels, (ii) **transparent**, specifying all involved semantic unit types, and (iii) **incrementally improvable**, supporting the identification of information gaps specific to their workflows and particular needs, and allowing improving the FAIRness incrementally in these directions, without having to reengineer entire datasets or data infrastructures.

With this chapter, we want to raise awareness for the need to rethink FAIRness and to develop new ways of assessing and representing differences in FAIRness across units of information of different granular richness and depth. Rather than asking *is this dataset FAIR?*, we should ask some other questions: **What parts of a dataset are FAIR, for whom, and why? What would it take to improve its FAIRness within a given use scenario? How should FAIRness scores propagate through nested modular data and knowledge structures? How should a FAIRness score relate to the granular complexity score of a dataset? How are cumulated FAIRness scores calculated for a dataset, if the FAIRness differs across its granularity levels? Is it possible to assess whether a given dataset is FAIR enough for a given task at hand (task-dependent FAIRness assessment)?**

All these represent open questions that have to be addressed at some point, given that we want to realise the IFDS. Such a perspective-sensitive model would integrate modularity, granularity, and user needs into FAIR assessment and would set the stage for the Grammar of FAIR and future FAIR and CLEAR infrastructures.

# Toward a Grammar of FAIR

The following presents our initial ideas for practical implementations, based on the analysis and conclusions above. These will be developed in more detail by a broader group of experts in the future.

## From Metaphor to Architecture

A natural language's grammar provides a generative and compositional framework for producing meaning. It includes a vocabulary (lexicon), rules for combining symbols (syntax and morphology), and systems of meaning (semantics). The **Grammar of FAIR** serves an analogous role for digital knowledge (Table 1): it provides a **semantic infrastructure** that enables the modular expression, combination, and reuse of machine-actionable meaning across information ecosystems.

Importantly, this grammar is not a syntactic coding scheme but a **semantics-first system**, grounded in cognitive linguistics and inspired by granularity, modularity, and evolvability. Its core units are **semantic** **units**, serialised as semantic FDOs and embedded within a layered system ensuring interoperability across formats, domains, and infrastructures.

Table 1: Mapping of various components and aspects of our Grammar of FAIR to a natural language grammar

| Language Component | Function | Grammar of FAIR Equivalent |
| --- | --- | --- |
| **Lexicon** | Vocabulary: inventory of words and terms | GUPRIs of concepts referred to by ontology terms and semantic unit types and their corresponding taxonomies |
| **Morphology** | Rules for forming words from roots, affixes, etc. | Ontologies that define atomic classes, properties, and relationships and thus foundational "morphemes" of FAIR information, from which more complex information structures (statements, schemata) are built |
| **Syntax** | Rules for combining words into phrases and sentences | Schema metamodels used in statement units, which provide the rules for combining ontology-defined entities into valid representations (e.g., SHACL shapes, SQL DDLs, JSON schemata). Additionally, from an information granularity perspective, also compound units that provide rules for combining statement units into semantically meaningful units of information of coarser levels of information granularity |
| **Phonology** | Sound patterns in spoken language | Dynamic labels and dynamic graphs |
| **Orthography** | Rules for writing and spelling in written language | Syntax constraints of file formats (RDF serialisation, JSON, CSV) |
| **Semantics** | Meaning of sentences and their components | Natural language token models and their schema-transitive representations that are based on formal semantics (i.e., a logical framework) |

These analogies are not merely illustrative, but express **structural constraints** that ensure semantic coherence and interoperability across human and machine users.

# Semantic Transitivity and Token Models

At the core of the Grammar of FAIR lies the principle of **semantic transitivity**: every semantic unit, regardless of format, must map to a **shared natural language token model**, which serves as a **semantic anchor** for deriving interoperable schema metamodels.

Take, for example, the sentence "*This sample was incubated at 37°C for 24 hours.*" It is a cognitively interoperable sentence, which can be translated across RDF/OWL, SQL, JSON, or any other technical implementation, as long as each representation is derived from the same language metamodel (e.g., "*ENTITY underwent PROCESS at CONDITION for DURATION*"; see Fig. 5). This language metamodel can be represented in a Rosetta Statement metamodel that can function as a machine-actionable reference model in the Grammar of FAIR. At the core of a granular and modular FDO construction is therefore that each and every concept in the structure (regardless of whether it functions in this structure as a subject, a predicate, or an object or any other role) should be referred to by a GUPRI, meaning a Globally Unique and Persistent Identifier which resolves to one (and only one) 'intended defined meaning'.

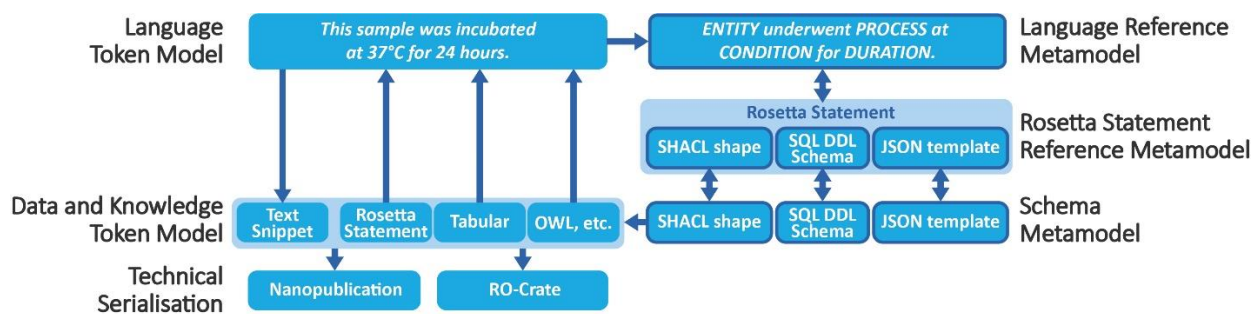

**Figure 5: Transitive architecture of the Grammar of FAIR.** Semantic content is expressed at different layers: 1) a natural language token model, representing a specific propositional instance (as dynamic labels or dynamic graph). 2) A natural language metamodel that is obtained from the language token model through classification and generalisation over its subject and object positions and that can be documented in a machine-actionable format as 3) a Rosetta Statement metamodel. The Rosetta Statement metamodel, in turn, functions as a reference metamodel for 4) one or more schema metamodels, each of which is linked to a particular semantic unit class and meets specific usage needs. Each schema metamodel is semantically interoperable with its Rosetta Statement reference metamodel using entity mappings and schema crosswalks, and each schema metamodel specifies 5) the structure of its corresponding data token models (semantic units). 6) The concrete technical serialisation of a data token model, depending on its technical specification as, for instance, a Nanopublication or an RO-Crate. This architecture ensures cognitive interoperability and format-agnostic machine-actionability through semantic transitivity.

This structure of the Grammar of FAIR enables seamless transformation without loss of meaning and mirrors the ***work-expression-manifestation*** distinction from FRBR, as the natural language token model and its corresponding language and Rosetta Statement metamodel corresponds to the ***work*** (semantic content), the data and knowledge token models and their corresponding schema metamodels are different ***expressions*** (formal realisations) of this *work*, and the technical serialisations (e.g., Nanopublications, RO-Crates, I-Adopt statements (59)) are ***manifestations*** (format-specific serialisations) of these *expressions*.

The Grammar of FAIR separates **semantic content from format**. Text snippets, RDF triples, SQL records, and Rosetta Statements are manifestations of the same content. What matters is **semantic coherence**, achieved through transitivity to a shared metamodel.

This decoupling enables cross-format transformations, FAIRness evaluation, and dynamic labelling across infrastructures. **Semantic units thus become the *lingua universalis***, i.e., a **foundational model**, of digital knowledge: modular, reusable, and translatable.

## FAIR and CLEAR Services as Grammatical Infrastructure

To operationalise the Grammar of FAIR, we require a suite of services that do more than deliver formats but embody FAIR and CLEAR principles, balancing **machine-actionability with human-actionability**, supporting governance, evolution, and semantic coherence (6,12):

1. **Terminology Service** (*Lexicon, Morphology*)
   - **Components**: Comprises a repository, registry and lookup service for terms defined in controlled vocabularies and ontologies, and for entity mappings of terms across different vocabularies for establishing terminological interoperability, providing the morphological and finest-grained information building blocks of semantic units. Each entity mapping is documented as an FDO that can be referenced via its own GUPRI and be curated independent of any specific vocabulary.
   - **Function**: Maintains and publishes canonical definitions of semantic unit classes, attributes, and relations, comparable to a lexicon authority.
   - **Role**: Supports consistent semantic annotation, term reuse, and disambiguation. Provides labels, definitions, multilingual text, synonyms, and entity mappings, all of which are both human- and machine-readable. With the entity mappings (e.g., via SSSOM (53)), it also supports ontology crosswalks, promoting reuse and interconnectivity.
   - **Impact**: Empowers semantic units to maintain typological consistency across domains.
   - **Governance**: Enables version control, stakeholder review, and growth pathways, per FAIR-services best practices.

2. **Schema Service** (*Syntax, Phonology*)
   - **Components**: Comprises a repository, registry, and lookup service for all kinds of schema metamodels (i.e., Rosetta Statement metamodels, SHACL shapes, SQL DDL schemata, JSON schemata, etc.) and schema crosswalks for establishing propositional interoperability. Each schema crosswalk is documented as an FDO that can be referenced via its own GUPRI and be curated independent of any specific vocabulary.
   - **Function**: Maintains alignment of schema metamodels with Rosetta Statement metamodels and thus language metamodels, providing dynamic labels and graphs for displaying the semantic content of a semantic unit in a cognitively interoperable format.
   - **Role**: Enables transitive schema validation, cross-format conversions, and compatibility traces in combination with respective operations/services from the Operations Service.
   - **Governance**: Tracks schema versions, dependencies, and change logs, which are essential for backward-compatible evolution.

3. **Operations Service**
   - **Components**: Comprises a repository, registry, and lookup service for all kinds of operations and data transformation and analysis related services, each aligned with compatible schema metamodels, for supporting AI-ready machine-actionability. Each operation/service is documented as an FDO that can be referenced via its own GUPRI and be curated independent of any specific vocabulary or schema metamodel.
   - **Function**: Provides a library of validated operations such as transformations, aggregations, queries, and logic-specific inference rules, and executable entity mappings and schema crosswalks, and from services that make these operations readily executable.
   - **Role**: Supplies machine-usable methods tied to schema metamodels and semantic unit types. This can be achieved by associating semantic unit classes via the GUPRIs of their schema metamodels to the GUPRIs of all operations/services that are interoperable with the schema.

4. **Workflow Service** (optional)
   - **Components**: Comprises a repository, registry, and lookup service for workflows consisting of chained-up individual operations. Each workflow is documented as an FDO that can be referenced via its own GUPRI.
   - **Function**: Enables orchestration of multi-step operations/services, such as a sequence of specific transformation, reasoning, and validation operations, over semantic units.
   - **Role**: Serves both machine pipelines and interactive data exploration (e.g., Jupyter, Rest APIs).
   - **Impact**: Making operations/services and workflows composable and referencable will increase the reproducibility of scientific findings and thus increase their trustworthiness.

These services each must follow (i) **FAIR metadata**, making their components discoverable, their usage restrictions clear via licensing, their provenance trails and versioning histories transparent, and (ii) **CLEAR criteria**, improving the cognitive interoperability of the services, their components, and the data and knowledge built by them, providing use-case-driven documentation, and improving overall transparency.

The FAIR Services scaffold the Grammar of FAIR, by providing the definitions of the vocabulary, structure, transformations, and governance of semantics. Crucially, they integrate human actionability (via documentation, dynamic labels and graphs, multilingual support) with machine pipelines, closing the semantic gap currently existing.

By combining terminology, ontology, schema, operations/services, and workflows under unified machine actionable governance, the service layer ensures that semantic units remain coherent, interoperable, and evolvable, thereby supporting decentralised infrastructures that can grow without losing meaning and interoperability.

## Toward Evolvable FAIRness

FAIR information ecosystems are not static and must evolve with the emergence of new knowledge, novel data types, logic formalisms, and usage needs. The Grammar of FAIR accommodates this by anchoring semantic units in natural language and enabling maintaining backward-compatible transitions via metamodel crosswalks.

Like biological systems, FAIR ecosystems evolve by **recombining stable, meaningful building blocks**. In this light, the Grammar of FAIR is not just a metaphor, but a **semantic constitution** for long-term knowledge stewardship and future information architectures. Some have called this *'a natural language for computers'*.

# Conclusion: From Modular Semantics to Scholarly Transformation

Semantic units, when implemented as semantic FDOs, do not merely structure data, information, and knowledge, they have the potential to reshape the architecture of scholarly communication. By modularising semantic content into independently identifiable, semantically meaningful units that are serialised into semantic FDOs, researchers gain the ability to **cite** not just publications, but **individual propositions**, **method statements**, or **data fragments**, each with their own persistent identifier and accompanying metadata.

This modular citation capacity, grounded in the **granularity of meaning** rather than monolithic documents, has transformative implications: **Authorship attribution** could shift from full-paper authorship lists to **statement-level provenance** (60); **citation practices** could evolve toward **qualified references** that support, contest, or contextualise specific claims and not just entire articles; and **academic reputation** could reflect **semantic contribution and impact**, rather than journal prestige or co-authorship inflation.

In such a scenario, a highly cited statement FDO, representing a specific measurement, insight, hypothesis, or claim, could become as career relevant as the supporting journal article. This would lower the entry threshold for making meaningful contributions, empower early-career researchers, and reduce the influence of commercial publishing venues in determining scholarly value.

This vision aligns with the broader trajectory of the IFDS when aligned with semantic units and enabled by FAIR and CLEAR services. In such an IFDS, each semantic FDO becomes not only a **first-class object** but a **node in a citation and knowledge network**. Researchers interact with data, information, and knowledge at multiple levels of granularity, creating, citing, transforming, and composing semantic units much like software engineers reuse modular code.

This is not an abstract future. It is a **converging path**, already visible in the architectures of Nanopublications, RO-Crates, schema registries, and the emerging tooling around FAIR and CLEAR principles. The Grammar of FAIR proposed in this article offers the epistemic, semantic, and technical scaffolding to build this future, grounded in logical structure, transitive models, and cognitive interoperability.

By aligning knowledge discovery and representation with modular semantics and FAIR- and CLEARness, we do not only manage complexity of information granularity; we **restructure how knowledge is created, shared, validated, and rewarded**.